\documentclass[aps,prl,floatfix,twocolumn,showpacs]{revtex4}

\usepackage{latexsym}
\usepackage{graphicx}

\usepackage{verbatim}
\usepackage{amsmath}
\usepackage{color}
\usepackage{comment}

\begin{document}

\title{Optoelectronic excitations and photovoltaic effect in strongly correlated materials}

\author{John E. Coulter$^{1}$}
\author{ Efstratios Manousakis$^{(1,2)}$ } \email{manousakis@magnet.fsu.edu}
\author{Adam Gali$^{(3,4)}$}
\affiliation{
$^{(1)}$ Department  of  Physics and National High Magnetic Field Laboratory,
  Florida  State  University,  Tallahassee,  FL  32306-4350,  USA\\
$^{(2)}$Department   of    Physics,   University    of   Athens,
  Panepistimioupolis, Zografos, 157 84 Athens, Greece \\
$^{(3)}$Institute for Solid State Physics and Optics, Wigner
  Research Center for Physics,\\
  Hungarian Academy of Sciences, P.O.B. 49, H-1525, Budapest, Hungary \\
$^{(4)}$Department of Atomic Physics, Budapest
              University of Technology and Economics, Budafoki \'ut 8., H-1111,
              Budapest, Hungary}

\date{\today}

\begin{abstract}
Solar cells based on conventional semiconductors have low efficiency
in converting solar energy into electricity because the excess energy 
beyond the gap of an incident solar photon
 is converted into heat by phonons. Here we show by \emph{ab initio} methods 
that the presence of strong Coulomb interactions in strongly 
correlated insulators (SCI) causes the highly photo-excited electron-hole pair 
to decay fast into multiple electron-hole pairs via impact ionization (II).
We show that the II rate in the insulating $M_1$ phase of vanadium dioxide
(chosen for this study as it is considered a prototypical SCI) is 
two orders of magnitude higher than in Si and 
much higher than the rate of hot electron/hole decay due to phonons. 
Our results indicate that a rather broad class of materials may be harnessed 
for an efficient solar-to-electrical energy conversion that has been not 
considered before.
\end{abstract}
\pacs{71.15.-m,71.15.Mb,78.,78.56.-a}
\maketitle

\begin{figure}
\includegraphics[width=\columnwidth]{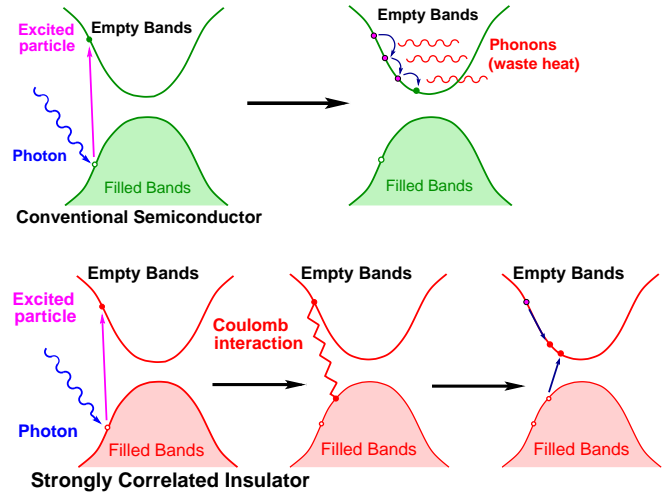}
\caption{The standard process in \emph{conventional semiconductors} is shown above (green bands) 
while the expected process in \emph{strongly correlated insulators} 
is shown below (red bands). See text for explanation. }
\label{fig:bands}
\end{figure}

Among several factors that practically limit solar cell efficiency 
is the fact that the excess energy of hot electrons, excited much above the semiconducting
gap by absorbing a high energy solar photon, is taken away by lattice
excitations within a time scale of the order of $10^{-13}-10^{-12}$ seconds
\cite{Si-phonon-Louie,Si-phonon-decay,Si-pump-probe-phonon}.
This is schematically illustrated  in Fig.~\ref{fig:bands} (top):
In \emph{ conventional semiconductors}
the incident solar photon promotes an electron
from the occupied valence band to the conduction band. The
excess energy of the excited ``hot'' electron beyond the
energy gap (the solar spectrum
ranges from 0.5~eV to 3.5~eV) is converted into waste heat by phonon emission 
and the
electron relaxes to its band edge within $10^{-13}-10^{-12}$ seconds.
If one chooses the material to have a large energy gap, 
solar photons with excess energy below the energy gap cannot excite electron
hole pairs, and all excess energy is lost to phonons.

There are many ideas on how to solve this problem (for 
example, see Ref.~\onlinecite{Nozik2002} and Ref.~\onlinecite{PbSe-IIR}).  
Following a recent suggestion\cite{Mottsolar}, 
we will demonstrate in this paper, by means of {\it ab initio} methods, that 
if strongly correlated insulators (SCI), such as the transition metal 
oxides (TMOs), are used as a basis for photovoltaic applications, one may see 
high quantum efficiency. As illustrated in Fig.~\ref{fig:bands} (bottom) in 
\emph{a strongly correlated insulator} 
the photo-excited electron (or hole) may utilize its strong Coulomb interaction 
(shown in figure by the red zig-zag line) with
another valence electron to promote it to the conduction band, thus,
creating a second electron-hole pair using the energy of the 
{\it same} solar photon. 
In a SCI, the localized electrons 
form an electronic system in which the residual effective 
electron-electron interaction 
is strong and can lead to a fast decay of the initially photo-excited 
electron/hole pair into multiple electron-hole pairs [so-called impact
ionization (II)] on a time-scale
much faster than other decay processes. 
 This clearly can lead to carrier 
multiplication, namely, the energy of a single incident photon is used
to create multiple electron-hole pairs instead of creating phonons.
This process, in Si and within the limits of the solar 
spectrum, takes place in about the same
time-scale as the phonon relaxation processes. 

Since an
insulator is needed for such photovoltaic applications,
it was further proposed\cite{Mottsolar} working at 
the simple model level
that a Mott insulator, where such strong correlations are present,
could be used. However, all that is needed for this idea to work is 
the strong Coulomb interaction and the essence of the proposal is still 
applicable in the more general case of strongly correlated materials, where the
insulating gap is not necessarily a Mott gap but of a some other origin;
for example, while VO$_2$ is a strongly correlated material, 
the gap is believed to be due to a Peierls instability.\cite{Coulter-scGW, PhysRevB.87.195106, PhysRevLett.108.256402} 
The possible increase of solar cell efficiency due to carrier multiplication
through the impact ionization
process has been proposed previously for traditional 
semiconductor nanocrystals.\cite{Nozik2002, PbSe-IIR} The first encouraging 
demonstration of the presence of multiple excitons in the photo-current
has been demonstrated in a photovoltaic cell based on semiconductor 
nanocrystals\cite{PbSe-Science-IIR}; however, the effect is not very strong.
Thus, alternative materials showing promise for carrier multiplication
upon solar illumination are still much sought-after.

In order to demonstrate that in SCI the II rate (IIR) is 
significantly enhanced we choose the room temperature phase of VO$_2$ 
(the $M_1$-phase) because VO$_2$ is considered  a prototypical strongly 
correlated insulator with a lot of history surrounding it.
We show that the residual electron-electron interaction in this material is
indeed very strong. Furthermore, we show that the IIR is 
much larger than the rate characterizing phonon processes and 
in the region of the solar
spectrum the IIR is orders of magnitude higher than that in Si.
VO$_2$ is used only to demonstrate increased IIR in strongly correlated 
materials, corresponding to multiple exciton generation (MEG).  It is not proposed 
as the optimal material for use in a photo-voltaic device due to other factors, 
some of which are discussed in our concluding remarks. 
The goal of the present paper is to
show that this more general class of strongly correlated insulators should be
explored experimentally to find the optimal material.
Therefore, here, we claim that strongly correlated materials  
can be good candidates to focus the search for efficient
photovoltaics. We particularly 
emphasize that a p-n junction has been very recently made\cite{VO2-device} 
from the insulating $M_1$ phase of 
VO$_2$ ($M_1$-VO$_2$). 

Multiple-carrier production due to impact ionization has
not been the subject of experimental investigation in this family 
of materials. It is of immediate 
importance to have a computational \emph{ab initio} scheme which is 
reliable to evaluate opto-electronic properties such as excitations, gaps, 
absorption, IIR and related properties 
for a given bulk or interface structure of TMOs. Many-body perturbation
 theory, such as the Bethe-Salpeter equation (BSE),
is useful for accurately calculating the optical properties of traditional
semiconductors.\cite{Rohlfing-BSE} However, it is an open question 
whether or not the BSE method based on well-established quasi-particle 
states and 
energies is able to provide qualitatively good results on the optical 
properties of the much less studied strongly correlated electron systems in 
TMOs such as $M_1$-VO$_2$. 

\begin{figure}
\includegraphics[width=\columnwidth]{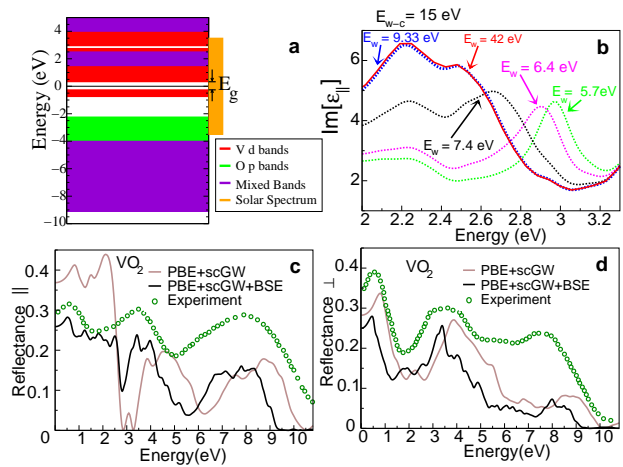} 
\caption{\textbf{a)} Schematic diagram on the bands of highly correlated electrons (vanadium d-orbitals) and conventional bands (s,p-orbitals). The origin is set in the middle of the fundamental band gap. $E_\text{g}$=0.63~eV is the calculated fundamental direct gap of $M_1$-VO$_2$. \textbf{b)} The convergence of the first absorption peak by the BSE method as a function of the valence band width ($E_w$ measured from the top of the valence band) included in 
the BSE calculation.
 \textbf{(c,d)} The experimental data taken from Ref.~\cite{PES} for the reflectance of $M_1$-VO$_2$ for polarization parallel (\textbf{c}) and perpendicular
  (\textbf{d}) directions, is compared with the results of PBE+scGW+BSE calculation. The PBE+scGW spectra are given to show the exciton effects captured by 
BSE method. The structure of $M_1$-VO$_2$ and the definition of directions are given in the Supplementary Material\cite{SuppMat}.}
\label{fig:bse-refl}
\end{figure}

As a first inevitable step in this direction, we study the optical 
properties of $M_1$-VO$_2$ by many-body perturbation techniques. 
We start with a self-consistent GW procedure (scGW)\cite{Hedin, 
Schilfgaarde} based on density functional theory in order to obtain 
quasi-particle energies and states (see Fig.~\ref{fig:bse-refl}a), which
has given good results on similar materials\cite{Svane2010}, including
strongly correlated f electron systems\cite{Chantis2007,Chantis2008}. 
This self-consistent procedure is the only \emph{ab-initio} way to obtain
consistent results, a good single particle gap and density of 
states for this system.\cite{Coulter-scGW} Then, we include the 
electron-hole interactions in quasi-particles via the BSE method
and calculate the optical properties of VO$_2$. 


We carry out density functional theory (DFT) calculations using the plane 
wave basis set (plane wave  cutoff of 400~eV) with the projector augmented 
wave (PAW) method \cite{Blochl}, as implemented within the \textsc{VASP} 
package.\cite{Kresse1, Kresse2, Kresse4, Kresse5} The experimental geometry of
$M_1$-VO$_2$ is used where the parallel direction of the lattice 
is defined along its $a$-axis according to the convention 
(see Supplementary for additional information). Either PBE\cite{PBE} or HSE06\cite{HSE06} functionals are 
applied as a starting 
point for the self-consistent GW procedure as described in 
Ref.~\onlinecite{Coulter-scGW}, in order to obtain quasi-particle energies 
\emph{and} states. We include the vanadium 3p electrons as valence in the 
calculations which results in 46 valence band (occupied) states that we 
label by $n_v$. In the GW procedure, we use 146 conduction band (virtual) 
states that we label by $n_c$. 
The inclusion of only $3p$ vs.\ $3s$ electrons was checked at the level of 
G$_0$W$_0$ and no difference was found within $\pm$10~eV of the
Fermi level.
Based on a recent publication\cite{VASP-GWbands}, one must
be very careful with the convergence of the plane-wave basis set in PAW-based
GW calculations.  While the absolute value of the energy eigenvalues
shows a significant change with number of bands and size of basis set,
the energy difference between eigenvalues is much less sensitive to these
parameters.
For example, the size of our plane-wave basis set ($\sim$850 plane waves) is
sufficient to produce a gap in the calculations
presented in Ref.~\onlinecite{VASP-GWbands} 
accurate to about 3\%.
Additionally, the number of conduction bands used here 
is sufficient to converge the band gap in Ref.~\onlinecite{VASP-GWbands}
to within 6\% of the value produced when the fully 
convergent limit of the number of bands is reached. This is impossible
to reach in the fully scGW scheme used here. Note that these
estimates are based on G$_0$W$_0$ calculations, and produce corrections
far smaller than those obtained with scGW.
For the optical properties we solve the BSE 
by diagonalizing the well-known electron-hole pair 
Hamiltonian\cite{Rohlfing-BSE}.
For VO$_2$, we used 26 occupied and unoccupied bands, on a 5$\times$5$\times$5 k-point 
mesh in the
BSE calculations presented here. For Silicon, we used 4 occupied 
and 8 unoccupied bands, on a 12$\times$12$\times$12 k-point mesh, with the experimental
geometry. We checked in detail that these sizes are 
large enough to produce converged results.  For details on convergence and 
implementation, please see the Supplementary Material\cite{SuppMat}.

By applying the above discussed implementation of the BSE method 
(see Supplementary\cite{SuppMat}) on top of scGW quasiparticle states
obtained in Ref.~\onlinecite{Coulter-scGW}, 
we are able to achieve excellent 
agreement with the experimental dielectric function\cite{Si-Expt} of 
Si (see Fig.~S3 of the supplementary material\cite{SuppMat}). This indicates that the current approach 
is a state-of-the-art {\it ab initio} method to 
be used for optical properties.

The major features in the calculated reflectance spectra of VO$_2$ broadly 
agree  with available experimental data for a wide range of 
excitation energies up to 10~eV:\cite{PES, Epsilon2data}  
The main peaks are accurately reproduced at about 
0.7~eV, 3.5~eV and 8.0~eV in the parallel direction and at about 0.7~eV, 
3.2~eV and 7.9~eV in the perpendicular direction 
(see Fig.~\ref{fig:bse-refl}). We note that our results are valid at $T$=0K 
but that the inclusion of the electron-phonon interaction, which might 
have a measurable effect in the optical spectrum observed at room 
temperature, is beyond the scope of the paper. Nevertheless, we conclude 
that the BSE method is able to provide good results on the absorption 
properties. Next, we focus our discussion on the threshold
of absorption (optical gap) which is the lowest energy excitation. 

The optical gap in the BSE calculation appears at 0.26~eV valid at $T$=0K. 
The absorption spectrum is recorded at room temperature, so that it is not 
obvious where the experimental onset of absorption is. Nevertheless, one may
observe that the absorption starts to rise at $\sim$0.4~eV in the bulk
\cite{Tomczak-2009, PES, Epsilon2data} while in thin films it rises at 
$\sim$0.5~eV.\cite{Tomczak-2009, Quazilbash-Expt, Okazaki-Expt} 
Photo-emission data indicates a fundamental single particle gap of $\sim$0.6~eV, 
in agreement with our scGW calculation.\cite{Coulter-scGW} 
Thus, the binding energy of the exciton is $\sim$0.2~eV. In the scGW 
calculation\cite{Coulter-scGW} the calculated fundamental 
direct gap is 0.63~eV, thus, the BSE exciton binding 
energy is $\sim$0.37~eV. Therefore, both 
the calculated optical gap and the binding 
energy of the exciton agree reasonably well with the experiment. 
We note that the optimal band gap of high IIR absorber materials for 
solar cell applications is about 0.8~eV
\cite{IIR-optimal-gap},
which further reduces toward $\sim$0.3~eV when solar concentrators are 
applied.\cite{IIR-concentrators}
This value lies close to the calculated optical gap of VO$_2$.

Next, we study the role of the electron-electron interaction by looking at 
the convergence of the optical gap as a 
function of the number of valence bands (see Fig.~\ref{fig:bse-refl}b). We find that the optical gap 
converges only when we include a large number of valence bands $n_v=26$ 
from the top of valence band where 
these occupied bands are $\sim$10~eV deep. This suggests that the 
Coulomb-interaction between the quasi-particles is very strong. We believe 
that this may be due to the
pronounced localized nature of the d$_{x^2-y^2}$ orbitals\cite{Coulter-scGW}
 and the strong electron correlation arising from the charge localization.
While this material is found to be a Peierls and not a Mott 
insulator\cite{PhysRevB.87.195106, PhysRevLett.108.256402}, the
conclusions of Ref.~\onlinecite{Mottsolar} are still applicable
here: the screened Coulomb interaction between electrons and holes,
which is found to be very strong in $M_1$-VO$_2$, might
cause a high impact ionization rate, leading to multi-exciton generation upon
high-energy excitation as explained in Fig.~\ref{fig:bands}.

In order to demonstrate that this large effective
Coulomb interaction can play a significant role in enhancing 
the IIR, next, we present our results for the IIR on Si and VO$_2$. 
An estimate of the IIR can be obtained  using the self-energy 
calculated from the scGW 
wave functions, similar to earlier work using G$_0$W$_0$.\cite{Kotani-IIR}
The IIR $\tau_{\vec{k} n}^{-1}$ is calculated from the self-energy as:
\begin{equation}
\tau^{-1}_{\vec{k} n} = \frac{2 Z_{\vec{k} n}}{\hbar} 
                        | \text{Im} \, \Sigma_{\vec{k} n} | \,;\,
Z_{\vec{k} n} = \left(1 - \text{Re} 
                \frac{\partial \Sigma_{\vec{k} n} (\omega)} 
                {\partial \omega}\Big|_{\epsilon_{\vec{k} n}} \right )^{-1}
\end{equation}
where $\vec{k}$ and $n$ are k-point and band indices. When $n$ band indices are running for valence bands then they correspond to the recombination to the hole-initiated biexcitons, otherwise to the electron-initiated biexcitons.\cite{Govoni2012}
For the self-energy calculations, we use the same parameters as in the 
scGW calculations described earlier. Further details on the calculation 
are given in the Supplementary Materials\cite{SuppMat}.

\begin{figure}
\includegraphics[width=0.9\columnwidth]{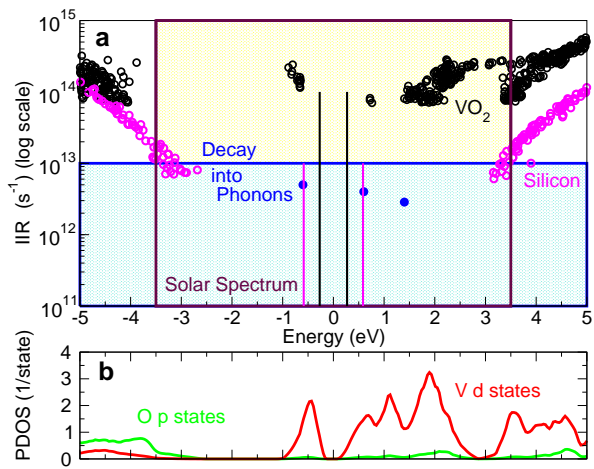}
\caption{
\textbf{a)} The IIR of Si and $M_1$-VO$_2$ within our
method. 
 Additionally, the regions of the solar spectrum and rate of decay 
into phonon modes\cite{Si-phonon-decay} are outlined for clarity. Points
where the decay rate of electrons to phonons has recently been 
calculated \cite{Si-phonon-Louie} are highlighted in blue.
The data points in Si and in VO$_2$ for IIR below certain cutoffs
are not sufficiently accurate (for details see Supplementary 
material\cite{SuppMat}), thus, we do 
not show those points here. The band gaps for Si and VO$_2$ are indicated
by vertical lines at approximately 1.2 and 0.6 eV respectively.
 \textbf{b)} Decomposition of 
IIR to the corresponding bands of $M_1$-VO$_2$. The highly correlated bands are
labeled by red curves originated from vanadium d-orbitals while the 
conventional bands in green are from the oxygen p-orbitals.}
\label{fig:IIR}
\end{figure}

We can give an indication that the IIR for VO$_2$ is much higher than
that of Si within the solar spectrum by comparing our results
on Si and VO$_2$.
As can be seen in Fig.~\ref{fig:IIR}, the IIR on VO$_2$ within
the region of the solar spectrum is  at least two orders of magnitude higher
than that on Si. Additionally we find, in agreement with the 
prediction,\cite{Mottsolar} that the IIR for VO$_2$ is significantly higher
(2-3 orders of magnitude)
than the rate of decay into phonons, which is required in order for
the impact ionization to be effective relative to phonon 
processes. Particularly, the hole-initiated biexcitons\cite{Govoni2012} 
associated with the sub-bands of 
d-electrons have an increased IIR while the p-bands show an IIR similar
 to the traditional Si semiconductor bands. This clearly demonstrates
that the highly correlated electrons indeed provide a high impact 
ionization rate which is the basis of multiple exciton generation upon 
solar illumination.
Here, we demonstrated that effective carrier multiplication should occur in
$M_1$-VO$_2$. We conclude that strongly correlated and particularly,
related TMO materials, can be very promising for
probing enhanced multiple exciton generation in 3D solids.

 A recent measurement on mono-crystalline monobeams of VO$_2$ gives us hope 
for efficient separation of photo-excited carriers in SCI.  Contrary to expectations, long recombination lifetime   of the order of 
microseconds upon laser illumination was found of the extracted
carriers.\cite{Miller12}
This surprisingly slow carrier recombination 
may be understood by the fact that certain optical transitions 
across the d-sub-bands of VO$_2$ are forbidden. We also find that typical 
velocities of electrons or holes of the VO$_2$ bands near the Fermi surface
are very high, a fact which when combined with the high VO$_2$ dielectric 
constant leads to large separation distances between the photo-excited 
electron-hole pairs within very short time scales. However,
VO$_2$ may not be the appropriate material to use for photovoltaic
applications for other reasons. For example, 
upon moderate heating (which occurs naturally upon solar illumination) 
it undergoes a metal insulator transition. In addition, VO$_2$ appears to
be ``self-doped'' with a relatively high n-type carrier density in the
pure gapped $M_1$ phase. However, the main point 
of the present paper is not to promote VO$_2$ as the material of choice for
photovoltaic applications. We have used VO$_2$ simply because it is a prototypical
material where strong electron correlations play an important role and we 
showed that the IIR is significantly increased.
The high IIR which is confirmed, here, to occur in 
SCI at such low energy in the solar 
spectrum gives solid indication that the multiple-carrier excitation effect would
be important when SCI are used
as basis for solar cells.\cite{Mottsolar} From within this broad class of materials one should 
select for further investigation 
those which exhibit other properties which are promising for
photovoltaic applications, including long carrier lifetime, small recombination 
rate\cite{Miller12} and enhanced mobility.


This work was supported in part by the U.S.\ National High Magnetic 
Field Laboratory, which is partially funded by the U.S.\ National 
Science Foundation. AG acknowledges the support from the Lend\"ulet programme of the Hungarian Academy of Sciences.



\end{document}